\documentclass[a4paper,reqno,11pt]{amsart}
\usepackage{amsmath,amssymb,amsthm}
\usepackage{cite}

\numberwithin{equation}{section}

\def\CS{\mathcal{S}}
\def\CP{\mathcal{P}}
\def\CT{\mathcal{T}}

\def\be{\beta}
\def\al{\alpha}

\def\tilde{\widetilde}

\def\ri{{\rm{i}}}

\def\?{(?)\marginpar{|?}}

\def\beq{\begin{equation}}
\def\eeq{\end{equation}}
\def\be{\begin{equation*}}
\def\ee{\end{equation*}}
\def\bea{\begin{eqnarray}}
\def\eea{\end{eqnarray}}

\begin{document}


\title[Multiscale expansion of the lattice potential KdV equation]
{Multiscale expansion of the lattice potential KdV equation on functions of infinite  slow-varyness order} 

\author{R. HERNANDEZ HEREDERO}

\address{Universidad Polit\'ecnica de Madrid\\ 
Escuela Universitaria de Ingenier\'ia T\'ecnica de Telecomunicaci\'on\\
Departamento de Matem\'atica Aplicada\\
Campus Sur Ctra de Valencia Km.~728031, Madrid, Spain}
\email{rafahh@euitt.upm.es}

\author{D. LEVI}

\address{Dipartimento di Ingegneria Elettronica \\
Universit\`a degli Studi Roma Tre and Sezione INFN, Roma Tre \\
Via della Vasca Navale 84, 00146 Roma, Italy}
\email{levi@fis.uniroma3.it}

\author{M. PETRERA} 

\address{Zentrum Mathematik \\
Technische Universit\"at M\"unchen \\
Boltzmannstr.~3, D-85747 Garching bei M\"unchen, Germany\\
and Dipartimento di Ingegneria Elettronica \\
Universit\`a degli Studi Roma Tre\\
Via della Vasca Navale 84, 00146 Roma, Italy}
\email{petrera@ma.tum.de}

\author{C. SCIMITERNA}

\address{Dipartimento di Fisica e Dipartimento di Ingegneria Elettronica\\
Universit\`a degli Studi Roma Tre and Sezione INFN, Roma Tre \\
Via della Vasca Navale 84, 00146 Roma, Italy}
\email{scimiterna@fis.uniroma3.it}


\begin{abstract}
We present a discrete multiscale expansion  of the lattice potential 
Korteweg-de Vries (lpKdV) equation on functions of infinite order of slow-varyness. 
To do so we introduce a formal expansion of the shift operator on many lattices holding at all orders. 
The lowest secularity condition from the expansion of the lpKdV equation gives a nonlinear lattice equation,
depending on shifts of all orders, of the form of the nonlinear Schr\"odinger (NLS) equation. 
\end{abstract}


\maketitle
\section{Introduction }

Reductive perturbation techniques~\cite{t1,t2} have proved to be important tools to find approximate solutions 
for many  physical problems by reducing a given nonlinear partial differential equation 
to a simpler equation, which is  often integrable~\cite{ce}. Recently a few attempts to carry over this approach  to partial difference equations
have been proposed~\cite{Ag,lm,LevHer,levi,lp}. 

The basic tool of the discrete reductive perturbation technique developed in~\cite{levi,lp} is a proper
multiscale expansion carried out by introducing various scales and lattices defined on them. 
Let us recall some of the main results. The transformation between two different lattices of indices~$n$ and~$n_1$ is given by~\cite{Jordan}
\beq
\Delta^j_{n}  u_{n}\doteq \sum_{i=0}^{j}  (-1)^{j-i}
{j \choose i} u_{n+i} = j! \sum_{i=j}^\infty  \frac{P_{i,j}}{ i!} \Delta^i_{n_1}  u_{n_1},  \label{difInt}
\eeq
where~$u_{n}: \mathbb{Z}\rightarrow \mathbb{R}$ is a function
defined on a lattice of index~$n \in \mathbb{Z}$ and~$u_{n_1}: \mathbb{Z}\rightarrow \mathbb{R}$ is
the same function on a lattice of index~$n_1 \in \mathbb{Z}$. By the symbol~$\Delta$ we mean 
the standard forward difference of the function~$u$ w.r.t.~its subscript, e.g.
$\Delta_n u_n \doteq u_{n+1} - u_n$. Here  the coefficients
$P_{i,j}$ are given by
$$
P_{i,j} \doteq \sum_{k=j}^i
 \omega^k \CS_{i}^{k}  \mathfrak{S}_{k}^j, 
$$
where~$\omega$ is the ratio of the increment in the lattice of variable~$n_1$ 
with respect to that of variable~$n$.
The coefficients~$\CS_{i}^{k}$ and~$\mathfrak{S}_{k}^j$ are the Stirling numbers of the
first and second kind respectively~\cite{AS}. Eq.~(\ref{difInt}) implies that a finite difference 
in the discrete variable~$n$ depends on an {\it infinite } number of differences on the variable 
$n_1$, e.g.~the function~$u_{n+1}$ can be written as a combination of the functions~$u_{i}$'s for~$i$ varying on an infinite 
subset of the lattice with index~$n_1$.   Formula (\ref{difInt}) can be inverted yielding
$$
\Delta^j_{n_1} u_{n_1}= j! \sum_{i=j}^\infty  \frac{Q_{i,j}}{i!}\Delta^i_{n}  u_{n},
\qquad \qquad 
Q_{i,j} \doteq \sum_{k=j}^i
 \omega^{-k} \CS_{i}^{k}  \mathfrak{S}_{k}^j.
$$

In~\cite{levi,lp} one has introduced the notion of {\it slow-varyness  of order~$\ell$}  for a function~$u_n$ iff~$\Delta_n^{\ell+1}  u_n=0$
(or equivalently~$\Delta_{n_1}^{\ell+1}  u_{n_1}=0$, see~\cite{lp} for further details).
This definition enables  to reduce infinite series to a finite number of terms.  Moreover one has considered
a generalization of formula (\ref{difInt}) in order to deal with functions
$u_n = u_{n; \{n_i\}_{i=1}^K}$ depending on a finite
number~$K$ of  lattice variables~$n_i$, $1 \leq i \leq K$.  
The computations done in~\cite{levi} proved that an integrable lattice equation, as the lattice potential KdV (lpKdV) equation,
reduces to a  completely discrete nonlinear Schr\"odinger (dNLS) equation
of the form
\beq \label{dnlstt}
 {\rm{i}} (\phi_{n,m+1} - \phi_{n,m}) +
  c_1( \phi_{n+1,m}- 2  \phi_{n,m}  +
\phi_{n-1,m} ) + 
 c_2\phi_{n,m} |{\phi}_{n,m}|^2= 0, 
\eeq
being~$c_1,c_2$ two real coefficients. Let us stress the fact that the above dNLS equation has a completely local
nonlinear part. We refer to the
papers~\cite{levi,lp}  for further details. 
It has been proved by singularity confinement~\cite{Papageorgiou}  and algebraic entropy~\cite{Viallet}
that the constructed dNLS equation is not integrable.

In~\cite{ce} Calogero and Eckhaus have shown that a necessary condition for 
the integrability of a nonlinear partial differential equation is that its multiscale reduction be integrable.
Here, trying to find a lattice analogue of the Calogero-Eckhaus theorem, we  
extend the techniques developed in~\cite{levi,lp} to any order of slow-varyness.

In Section~\ref{sec1} we introduce a formal multiscale expansion holding at all orders of slow-varyness.
Then in Section~\ref{sec2} we apply this technique to the lpKdV equation, thus providing an extension to all orders of the dNLS equation obtained in~\cite{levi,lp}. 
Finally, Section~\ref{sec4} is devoted to concluding remarks.

\section{Multiscale expansion on a lattice} \label{sec1}

\subsection{Lattices and shifts defined on them}

Let~$u_n: \mathbb{Z}\rightarrow \mathbb{R}$ be a function defined on a lattice of index~$n\in\mathbb{Z}$. 
We can always extend it to a real function~$u: \mathbb{R} \rightarrow \mathbb{R}$ by 
defining a real continuous variable 
$x \doteq n \sigma_x$, where~$\sigma_x \in \mathbb{R}$ is the lattice spacing.

Let us define the shift operator~$T_n$
such that~$T_n u_n \doteq u_{n+1}$. For~$u(x)$ we can introduce the operator~$T_x$,
corresponding to~$T_n$,
such that~$T_x u(x)\doteq u(x+\sigma_x)$.
The Taylor expansion of~$u(x+\sigma_x)$ centered in~$x$ reads
\bea \label{tay1}
T_x u(x)=u(x)+\sigma_x u^{(1)}(x)+\frac{\sigma_x^2}{2
}u^{(2)}(x)+ ... +\frac{\sigma_x^i}{i!}u^{(i)}(x)+ ... = \sum_{i=0}^\infty \frac{\sigma_x^i}{i!} u^{(i)}(x),
\eea
where~$u^{(i)}(x) \doteq d^i u(x) /dx^i \doteq d_x^i u(x)$, being~$d_x$  the total derivative operator. Eq.~(\ref{tay1}) suggests  the following formal expansion for the differential operator~$T_x$:
\beq \nonumber
T_{x}\doteq e^{\sigma_x d_{x}}= \sum_{i=0}^\infty \frac{\sigma_x^i}{i!} d_x^i.
\eeq
Introducing a formal derivative with respect to the index~$n$, say~$\delta_n$, 
we can define, by analogy with~$T_x$, the operator~$T_n$  as
\beq \label{tnr}
T_{n} \doteq e^{\delta_{n}}= \sum_{i=0}^\infty \frac{\delta_n^i}{i!}.
\eeq
The formal expansion (\ref{tnr}) can be inverted, yielding
\beq \label{tnn}
\delta_{n}=\ln{T_{n}}=\ln(1+\Delta_{n})=  \sum_{i=1}^\infty \frac{(-1)^{i-1}}{i}\Delta_{n}^i,
\eeq
where~$\Delta_{n}\doteq \Delta_{n}^+ \doteq T_{n}-1$ is the discrete first right difference operator w.r.t.~the variable~$n$, see eq.~(\ref{difInt}).
Notice that this is just one of the possible inversion formulas for the operator~$\delta_{n}$. For example it can be also written
in terms of left difference operators~$ \Delta_{n}^- \doteq 1 -T_{n}^{-1}$:
$$
\delta_{n}=- \ln{T_{n}^{-1}}=-\ln(1-\Delta_{n}^-)=  \sum_{i=1}^\infty \frac{(\Delta_{n}^-)^i}{i},
$$
or in terms of symmetric difference operators~$ \Delta_{n}^s \doteq (T_n -T_{n}^{-1})/2$:
$$
\delta_{n}= {\rm{sinh}}^{-1} \Delta_{n}^s=  \sum_{i=1}^\infty \frac{P_{i-1}(0)}{i} (\Delta_{n}^s)^i,
$$
where~$P_i(x)$ is the~$i$-th Legendre polynomial evaluated in~$x=0$.
Hence the~$\delta_n$
operators are formal series containing infinite
powers of~$\Delta_n$, but, acting on slow-varying functions of order~$\ell$, they reduce
to polynomials in~$\Delta_n$ of order at most~$\ell$. Consequently, any formula written
in terms of powers of~$\delta_n$ for a given slow-varyness order~$\ell$, contains its version for
smaller orders~$j<\ell$. 

The convergence of the series~$\delta_{n} u_n$ depends on the
analyticity properties of the function~$u_n$. Hence,
from now on, we will proceed formally, considering that
the developments can be justified a posteriori, because the functions which
finally appear (solutions to the resulting  difference
equations)  will have the correct analyticity properties.

\subsection{Dilations on the lattice}

Let us introduce a second lattice, obtained from the first one by 
a dilation. At first it is convenient to visualize the problem as a
change of variable between the continuous variable~$x \in\mathbb{R}$  and a new
continuous variable~$x_1\doteq {\epsilon}{x}$, $0 < \epsilon \ll 1$. 
On the lattice one considers a change from the index~$n\doteq x/ \sigma_{x}$ 
to the new index~$n_1 \doteq x_1/ \sigma_{x_1}$, 
where~$\sigma_{x_1}$ is the new spacing.
 Assuming that~$\sigma_{x_1} \gg \sigma_x $ we
can set~$\sigma_x \doteq \varepsilon \sigma_{x_1}$, $0 < \varepsilon \ll 1$, so that~$n_1 = \epsilon \varepsilon n$.
As~$n,n_1\in \mathbb{Z}$, $\epsilon \varepsilon$ is a rational number and one can define in all generality
$\epsilon \varepsilon\doteq  M_1/N \ll 1$ with~$M_1,N\in\mathbb{N}$. However, if we want that the lattice of index~$n_1$ is a sublattice of the lattice of index~$n$, we have also to require that~$M_1/N  =1/M$ with~$M\in\mathbb{N}$.

The relationship between the discrete
derivatives defined in the two lattices
is given by eq.~(\ref{difInt}), for which we shall give a straightforward proof based on the  well-known formulas~\cite{AS}
$$
( e^x-1)^j=j!\sum_{k=j}^{\infty}\frac{\mathfrak{S}_k^{j}}{k!} x^k, 
\qquad \qquad
[\ln(1+x)]^k=k!\sum_{i=k}^{\infty} \frac{\mathcal{S}_i^k} {i!} x^i.
$$
Here~$n_1= n (M_1/N)$ and thus~$\omega=M_1 /N$.
From the expression of~$\Delta_{n}^j$, $j \in \mathbb{N}$, we get by a straightforward algebra:
\bea
\Delta_{n}^j&=&(T_n-1)^j =(e^{\sigma_xd_x}-1)^j= 
j!\sum_{k=j}^{\infty}\mathfrak{S}_k^{j}\frac{(\sigma_x d_x)^
k}{k!}= \nonumber \\
&=&  j!\sum_{k=j}^{\infty}\frac{\mathfrak{S}_k^{j}}{k!}\left(\frac{\epsilon \sigma_x}{
\sigma_{x_1}}\right)^k(\sigma_{x_1} d_{x_1})^k=
j!\sum_{k=j}^{\infty}\frac{\mathfrak{S}_k^{j}}{k!}\left(\frac{\epsilon \sigma_x}
{\sigma_{x_1}}\right)^k[\ln(1+\Delta_{n_1})]^k = \nonumber \\
&=& j!\sum_{k=j}^{\infty}\frac{\mathfrak{S}_k^{j}}{k!}\left(\frac{\epsilon \sigma_x}{
\sigma_{x_1}}\right)^k
k!\sum_{i=k}^{\infty} \frac{\mathcal{S}_i^k}{i!}\Delta_{n_1}^i=j!
\sum_{i=j}^{\infty}\frac{1}{i!}\left[\sum_{k=j}^{i}\left(\frac{\epsilon
\sigma_x}{\sigma_{x_1}}\right)^k\mathcal{S}_i^k\mathfrak{S}_k^{j}
\right]\Delta_{n_1}^i. \nonumber
\eea
As~$\sigma_{x}=\varepsilon \sigma_{x_1}$ and~$\epsilon \varepsilon =  M_1/N$ eq.~(\ref{difInt}) is proven.

\subsection{Discrete multiscale expansion}

We present here the formulas necessary to construct a discrete multiscale expansion.
According to the definitions given above, let us consider~$u \doteq u_{n;n_1} = u(x; x_1)$ 
as a function depending on a fast index~$n$ 
and a slow index~$n_1= n( M_1/N)$. 
At the continuous level, the total
derivative~$d_{x}$ acting on functions~$u(x;x_1)$ is the sum of partial
derivatives, i.e.~$d_{x}=\partial_{x} +\epsilon\partial_{x_1}$. 
What is the situation on the lattice? Let us construct a relation
between the
total shift operator~$T_{n}$ and the partial shift
operators~$\CT_n,\CT_{n_1}$ ($\mathcal{T}_{n}u_{n;n_1} = u_{n+1;n_1},\mathcal{T}_{n_1}u_{n;n_1} = u_{n;n_1+1}$).
As
$$
T_{x}= e^{\sigma_x d_{x}}=
e^{\sigma_x \partial_{x}} e^{\epsilon \sigma_x \partial_{x_1}},
$$
we can write
\beq
T_n =
e^{\delta_{n}}
e^{ (M_1/N)\delta_{n_1}}\doteq\mathcal{T}_{n} \mathcal{T}_{n_1}^{(  M_1/N)}, \label{ll}
\eeq
with
\beq
\mathcal{T}_{n} \doteq  \sum_{i=0}^\infty \frac{\delta_n^i}{i!}, \qquad \qquad 
\mathcal{T}_{n_1}^{( M_1/N)}\doteq
\sum_{i=0}^{\infty}\frac{( M_1/N)^{i}}{i!}\delta_{n_1}^i, \label{deftxi}
\eeq
where~$\delta_{n_1}$ is given by eq.~(\ref{tnn}) with~$n$ substituted by~$n_1$.

Eq.~(\ref{ll}) can be easily extended to the case of~$K$
slow variables~$x_{i}\doteq \epsilon^{i}x$, $1 \leq i \leq K$ . The action of the shift operator~$T_n$  on a
function~$u\doteq u_{n;\{n_i\}_{i=1}^K}$ depending on both  fast and slow variables can be written in terms 
of the partial shifts~$\mathcal{T}_{n},\mathcal{T}_{n_i}$:
\beq \label{txu}
T_n \doteq \mathcal{T}_{n}\prod_{i=1}^K \mathcal{T}_{n_{i}}^{(\epsilon_{n_i})} ,
\eeq
where the~${\epsilon_{n_i}}$'s are suitable functions of~$\epsilon$ and~$\varepsilon$ depending parametrically
on some integer coefficients~$M_i \in \mathbb{N}$, $1 \leq i \leq K$.

To develop the  fields appearing in partial difference equations with two independent discrete variables
one has to consider the action of the operator (\ref{txu}) on a function depending on two fast indices~$n$ and~$m$, and on  a set of~$K_n+K_m $ slow variables~$\{n_i\}_{i=1}^{K_n}$ and~$\{m_i\}_{i=1}^{K_m}$, i.e.~on 
$u \doteq u_{n,m; \{n_i\}_{i=1}^{K_n},\{m_i\}_{i=1}^{K_m}}$.
Notice that in principle it is possible to consider~$K_n=K_m=\infty$.
For the moment we assume a common definition of the small parameter 
$\epsilon$ for both discrete variables~$n$ and~$m$,  but 
we denote with~$M_i$ the integers for the slow variables~$n_i$ and with~$\tilde M_i$ the ones for~$m_i$. We have:
\bea  \nonumber
\epsilon_{n_i} \doteq  \frac{M_i}{N^{i}},  \quad 1 \leq i \leq K_n,
\qquad \qquad \epsilon_{m_i} \doteq \frac{\tilde M_i}{N^{i}}, \quad 1 \leq i \leq K_m.
\eea

In the next Section we shall consider a partial difference equation defined on a quadrilateral lattice, namely an
equation of the type~$f(u,T_n u, T_m u, T_n T_m u )=0$. Assuming~$K_n=1$ and~$K_m = K$
we get from eqs.~(\ref{deftxi},~\ref{txu}) the following expansions for the shift operators appearing in the equation
$f(u,T_n u, T_m u, T_n T_m u )=0$:
\beq
T_n = \mathcal{T}_{n}\mathcal{T}_{n_1}^{(\epsilon_{n_1})}=
\mathcal{T}_{n}
\left[1+ \frac{1}{N}  M_1\delta_{n_1}+\frac12 \frac{M_1^2}{N^2} \delta_{n_1}^2+ 
\frac16 \frac{M_{1}^3}{N^3}  \delta_{n_1}^3 +O(1/N^4)\right],\label{ttx} 
\eeq
\bea
T_m&=&\mathcal{T}_{m} \prod_{i=1}^K \mathcal{T}_{m_{i}}^{(\epsilon_{m_i})} = \label{ttt} \\
&=&\mathcal{T}_{m}
\left[1+ \frac{1}{N} \tilde M_1\delta_{m_1}+\frac{1}{N^2}\left(\frac{\tilde M_1^2}{2} \delta_{m_1}^2+\tilde M_{2}\delta_{m_2} \right) 
+ \right. \nonumber\\
&& \left.
\qquad + \frac{1}{N^3} \left(\frac{\tilde M_{1}^3}{6} \delta_{m_1}^3+
\tilde M_1 \tilde M_2 \delta_{m_1}\delta_{m_2}+\tilde M_{3}\delta_{m_{3}} \right) +O(1/N^4)
\right], \nonumber 
\eea
\bea
T_nT_m &=&
\mathcal{T}_n  \mathcal{T}_{n_1}^{(\epsilon_{n_1})} \mathcal{T}_m\prod_{i=1}^K \mathcal{T}_{m_{i}}^{(\epsilon_{m_i})}  = \label{tnm} \\
&=&\mathcal{T}_n\mathcal{T}_m\left[\vphantom{\frac{\epsilon_{t_1}^3}{2}} 1+\frac{1}{N} \left(M_1\delta_{n_1}+ \tilde M_1 \delta_{m_1}\right)+
\right. \nonumber 
 \\ \nonumber
 &&\qquad\left.
 +\frac{1}{N^2} \left(\frac{M_1^2}{2}\delta_{n_1}^2+ M_1 \tilde M_1 \delta_{n_1} \delta_{m_1}+\frac{ \tilde M_1^2}{2}\delta_{m_1}^2
+ \tilde M_2 \delta_{m_2}\right) + \right.
\\ \nonumber
&&\qquad\left. + \frac{1}{N^3} \left(\frac{M_1^3}{6}\delta_{n_1}^3
+\frac{M_1^2}{2} \tilde M_1\delta_{n_1}^2 \delta_{m_1}
+ \frac{\tilde M_1^2}{2} M_1 \delta_{n_1}\delta_{m_1}^2
+M_1 \tilde M_2 \delta_{n_1}\delta_{m_2} +
\right.\right.
\\ \nonumber
&&\qquad\qquad\left.\left.
+\frac{\tilde M_1^3}{6}\delta_{m_1}^3
+\tilde M_1 \tilde M_2 \delta_{m_1}\delta_{m_2}+\tilde M_3 \delta_{m_{3}}^3\vphantom{+\frac{\delta_{m_1}^2}{2}}\right)
+O(1/N^4)
\right].
\eea

\section{Multiscale expansion of the lpKdV equation} \label{sec2}

The lattice potential Korteweg-de Vries (lpKdV) equation is given by~\cite{frank}:
\bea
\CP \doteq (p-q + u_{n,m+1} - u_{n+1,m})(p+q - u_{n+1,m+1} + u_{n,m})-(p^2-q^2) = 0,\label{kdv}
\eea
where~$p,q$, $p\neq q$, are two real parameters.
The above equation is probably the best-known completely discrete nonlinear equation which involves just four 
points which lay on two orthogonal infinite lattices
and it is nothing else but the nonlinear superposition formula for the Korteweg-de Vries equation. 

By defining~$\mu\doteq p-q$ and~$\zeta\doteq p+q$, eq.~(\ref{kdv}) can be written as 
\begin{equation}\label{eqdKdV}
\CP \doteq  \left[ \mu(T_{n}T_{m}u-u)+\zeta(T_{n}u-T_{m}u)\right]-
\left[(T_{n}u-T_{m}u)(T_{n}T_{m}u-u)\right] \doteq \CP_\ell-\CP_{n\ell}=0,
\end{equation}
where~$\CP_\ell$ and~$\CP_{n\ell}$ denote respectively the linear and the nonlinear part of the lpKdV equation.
The linear part~$\CP_\ell$ has a travelling wave solution of the form
$u=\exp{\{\ri[ \kappa n -\omega(\kappa)m]\}}$ with
\begin{equation}\label{disp}
\omega(\kappa)=-{2}\arctan{\left(\frac{\zeta+\mu}{\zeta-\mu}\tan{\frac{\kappa }{2}}\right)}.
\end{equation}

Applying the expansions (\ref{ttx},~\ref{ttt},~\ref{tnm}) to the function~$u\doteq u_{n,m; n_1, \{m_i\}_{i=1}^K}$, $\CP_\ell$ takes the form
 \beq
\CP_\ell = \sum_{i=0}^\infty  \frac{1}{N^i} L_{i}   u,
\nonumber
\eeq
where the operators 
$L_i \doteq L_i(\mathcal{T}_{n},\mathcal{T}_{m}, \delta_{n_1} ,\{ \delta_{m_j}\}_{j=1}^K )$
can be constructed in a recursive way.
The lowest operators~$L_i$, $i=0,1,2$, read
\bea
L_0&\doteq&\mu(\CT_{n}\CT_{m}-1)+\zeta(\CT_{n}-\CT_{m}), \nonumber \\
L_1&\doteq& \mu \mathcal{T}_{n}\mathcal{T}_m\left(M_1\delta_{n_1}
   +\tilde M_1\delta_{m_{1}}\right)+\zeta\left(
  M_1 \mathcal{T}_n \delta_{n_1}- \tilde M_1 \mathcal{T}_m \delta_{m_1}\right) , \nonumber \\
L_2&\doteq& \mu \mathcal{T}_{n}\mathcal{T}_m\left[\frac{M_1^2
   }{2}\delta_{n_1}^2+M_1 \tilde M_1 \delta_{n_1}\delta_{m_{1}}+\frac{\tilde M_1^2}{2}\delta_{m_1}^2+\tilde M_2 \delta_{m_2}\right]+
  \nonumber \\
&& + \; \zeta\left[ \frac{M_1^2}{2} \mathcal{T}_n  \delta_{n_1}^2 -
   \mathcal{T}_m\left(\frac{\tilde M_1^2}{2}\delta_{m_1}^2 +\tilde M_2\delta_{m_2}\right)\right]. \nonumber 
\eea

As nonlinearity generate harmonics, let us now expand the function~$u$ as
\beq 
\nonumber
u \doteq\sum_{\alpha\in\mathbb{Z}}u^{(\alpha)}(n_1, \{m_i\}_{i=1}^K; N)
e^{ \ri \alpha(\kappa
n-\omega m)}.
\eeq
As~$u$ is assumed to be real then~$u^{(-\alpha)}=\bar u^{(\alpha)}$, 
where by~$\bar u$ we denote the complex conjugate of~$u$. 
Moreover, if the nonlinear part should enter as a perturbation in the multiscale expansion of eq.~(\ref{eqdKdV}), 
we need~$u^{(\alpha)}(n_1, \{m_i\}_{i=1}^K; \infty) = 0$, $\forall \al \in \mathbb{Z}$. 
This implies that we have to expand  each function~$u^{(\alpha)}$ in inverse powers of~$N$:
\begin{equation} 
\nonumber
u^{(\alpha)}(n_1, \{m_i\}_{i=1}^K; N)
\doteq \sum_{k=1}^{\infty} \frac{1}{N^k}
u^{(\alpha)}_k (n_1, \{m_i\}_{i=1}^K).
\end{equation}
Then~$\CP_\ell$ reads
\beq
\CP_\ell = \sum_{\alpha\in\mathbb{Z}}
\sum_{i=1}^\infty  \frac{1}{N^i} \sum_{k=0}^{i-1} L_{k} 
u^{(\alpha)}_{i-k} (n_1, \{m_j\}_{j=1}^K) e^{\ri \alpha ( \kappa n - \omega m)}. \nonumber
\eeq

Performing the multiscale expansion of eq.~(\ref{eqdKdV})
we get several  determining equations obtained selecting the different powers of 
$1/N$ and the different harmonics~$\alpha$. 

So, let us  write down the resulting determining equations at the lower orders of~$1/N$ 
for the harmonics~$\alpha=0,1,2$,  necessary 
to get a dNLS equation as a secularity condition. 

The order~$1/N$ gives, for~$\alpha=0, 1$,  linear equations
which are identically satisfied by taking into account the dispersion relation  (\ref{disp}). For~$ |\alpha| \geq 2$
one gets some linear equations whose only solution is given by~$u^{(\alpha)}_{1}=0$.

The order~$1/N^2$ gives, for the harmonics~$\alpha=0,1,2$,  the following equations
\bea
&
\left[ (\mu+\zeta) M_1 \delta_{n_1}+(\mu-\zeta)\tilde M_1 \delta_{m_1} \right]
 u_1^{(0 )} =2 \left(-e^{\ri \kappa}- e^{-\ri \kappa} +e^{\ri \omega } + e^{-\ri \omega }\right) |u_{1}^{(1)}|^2, \label{equ10}\\
& 
e^{\ri  \kappa } \left( \mu e^{ -\ri \omega } + \zeta \right)M_1 \delta_{n_1}
u_1^{(1 )}+ e^{- \ri \omega}
\left(\mu e^{\ri \kappa }-\zeta  \right)\tilde M_1 \delta_{m_1}
 u_1^{(1)}=0, \label{equ11}\\
& \left[\zeta  \left(e^{2 \ri \kappa }-e^{-2\ri \omega }\right)+ \mu\left(e^{2 \ri( \kappa -\omega)}-1\right)
   \right]u_2^{(2)}
   =\left(-e^{\ri \kappa }+e^{-\ri \omega }+e^{ \ri (2\kappa -\omega)  }-e^{\ri(
   \kappa  -2 \omega)  }\right) (u_1^{(1)})^2. \label{equ22}
\eea
The solution of eq.~(\ref{equ11}) is given by~$u_1^{(1)}(n_1,\{m_i\}_{i=1}^K)=u_1^{(1)}(n_2,\{m_i\}_{i=2}^K)$ with~$n_2 \doteq n_1+ \gamma m_1$, $\gamma \doteq \mp 1$, provided that the integers~$M_1$ and~$\tilde M_1$
are chosen as
\beq \label{MMw}
M_1= \gamma S e^{- \ri \omega} \left( \mu e^{ \ri \kappa }-\zeta \right), \qquad \qquad
\tilde M_1= - S e^{\ri  \kappa } \left( \mu e^{ -\ri \omega } + \zeta \right),
\eeq
where~$S \in \mathbb{C}$ is a constant. As it has been shown in~\cite{lp} one can always choose~$S= r \exp{(\ri \theta)}$,
with~$r>0$ and~$\theta= -\arctan \left[ (\zeta \sin \kappa) /(\zeta \cos\kappa -\mu) \right]$,  in such a way that~$M_1$ and
$\tilde M_1$ are indeed positive integers. Taking into account the dispersion relation (\ref{disp}),  the coefficients~$M_1$ and~$\tilde M_1$ in eq.~(\ref{MMw})  can be rewritten  as
\beq \label{MMw2}
M_1= \gamma S  \left( \mu - \zeta  e^{\ri  \kappa }  \right), \qquad \qquad
\tilde M_1= S e^{\ri  \kappa } \frac{\zeta^2 -\mu^2}{ \mu e^{\ri  \kappa }  -\zeta}.
\eeq

Eqs.~(\ref{equ10},~\ref{equ22}) allow to express~$u_1^{(0)}$ and~$u_2^{(2)}$ in terms of~$u_1^{(1)}$ and~$\bar u_1^{(1)}$.
As~$u_1^{(1)}$ is a function of~$n_2$ the same must be for~$u_1^{(0)}$ and~$u_2^{(2)}$, i.e.~$u_1^{(0)}(n_1,\{m_i\}_{i=1}^K)=u_1^{(0)}(n_2,\{m_i\}_{i=2}^K)$, $u_2^{(2)}(n_1,\{m_i\}_{i=1}^K)=u_2^{(2)}(n_2,\{m_i\}_{i=2}^K)$.
 Then, taking into account eq.~(\ref{MMw2}), we find that eqs.~(\ref{equ10},~\ref{equ22}) reduce respectively to
\bea 
& \delta_{n_2} u_1^{(0)}
= \alpha_1 |u_{1}^{(1)}|^2, \qquad \qquad &
\alpha_1 \doteq - \frac{2 \gamma \left(1+e^{\ri \kappa} \right)^2}
{ S e^{\ri \kappa} (\mu +\zeta) \left( \mu -\zeta e^{\ri \kappa} \right) },
\label{nn} \\ \label{equ22b}
& u_2^{(2)}
= \alpha_2 (u_{1}^{(1)})^2, \qquad \qquad & \alpha_2 \doteq \frac{1+e^{\ri \kappa}}
{(1-e^{\ri\kappa})(\mu+\zeta)}.
\eea

We can now consider the equation for  the harmonic~$\al=1$ at order~$1/N^3$. We have:
\bea
&&\left( \sigma_1 \delta_{n_1} +\sigma_2 \delta_{m_1} \right) u_2^{(1)}+
\left(  \sigma_3 \delta_{n_1}^2 
+ \sigma_4 \delta_{m_1}^2 + \sigma_5 \delta_{n_1} \delta_{m_1}
+\sigma_6 \delta_{m_2} \right) u_1^{(1)} = \label{nls} \\
&& \qquad = u_1^{(1)}   \left( \sigma_7 \delta_{n_1} +\sigma_8 \delta_{m_1}  \right) u_1^{(0)} +
\sigma_9 \bar u_1^{(1)} u_2^{(2)}, \nonumber
\eea
where, taking into account eq.~(\ref{MMw2}), the coefficients~$\sigma_i$, $1 \leq i \leq 9$, read
\bea
&&\sigma_1 \doteq \frac{\gamma S e^{\ri \kappa } \left( \mu - \zeta e^{\ri \kappa } \right)(\mu^2-\zeta^2)}{\mu e^{\ri \kappa } -\zeta}, 
\qquad  \qquad \qquad 
\sigma_2 \doteq -\gamma \sigma_1,\nonumber \\
&&\sigma_3 \doteq \frac12 \gamma S \left(\mu -\zeta e^{\ri \kappa }\right) \sigma_1  ,
\qquad  \qquad \qquad \qquad \quad \;\;\,
\sigma_4 \doteq  \frac{\sigma_1^2}{2 \left(\mu -\zeta e^{\ri \kappa }\right) }  ,\nonumber \\
&&\sigma_5 \doteq - \frac{\gamma \mu \sigma_1^2 }{\left(\mu^2 -\zeta^2\right)},  
\qquad  \qquad \qquad \qquad \qquad \qquad \; \;
\sigma_6 \doteq  \tilde M_2 \left(\mu  -\zeta e^{\ri \kappa}\right),\nonumber \\
&&\sigma_7 \doteq \frac{\sigma_1 \left(e^{2 \ri \kappa}-1\right)}{e^{\ri \kappa}(\mu+\zeta)}, 
\qquad  \qquad \qquad \qquad \qquad \qquad 
\sigma_8 \doteq \frac{S e^{\ri \kappa} \left( \mu^2-\zeta^2\right) (\mu+\zeta ) \left(1 - e^{2 \ri \kappa}\right)}{\left(\mu e^{\ri \kappa} -\zeta \right)^2}, \nonumber 
\eea
$$
\sigma_9  \doteq \frac{\zeta \mu \left( e^{2\ri \kappa} -1\right)^2 \left( e^{\ri \kappa} +1 \right)^2
(\mu-\zeta)}{e^{\ri \kappa} \left(\mu - \zeta e^{\ri \kappa}\right) \left( \mu e^{\ri \kappa} -\zeta\right)^2}  .\nonumber 
$$

Using eqs.~(\ref{nn},~\ref{equ22b}) we can write eq.~(\ref{nls}) as
\beq
\left( \sigma_1 \delta_{n_1} +\sigma_2 \delta_{m_1} \right) u_2^{(1)} = \mathcal{L}(u_1^{(1)}),\label{jh}
\eeq
where~$\mathcal{L}$ is a linear operator. Notice that the l.h.s.~of eq.~(\ref{jh}) is the same  as in formula (\ref{equ11}),
but it involves the field~$u_2^{(1)}$, instead of~$u_1^{(1)}$.
Requiring that no secular term appears we get  the following equation for~$u_1^{(1)}=u_1^{(1)}(n_2,\{m_i\}_{i=2}^K)$:
\beq
\left(  \sigma_3 \delta_{n_1}^2 
+ \sigma_4 \delta_{m_1}^2 + \sigma_5 \delta_{n_1} \delta_{m_1}
+\sigma_6 \delta_{m_2} \right) u_1^{(1)} = u_1^{(1)}   \left( \sigma_7 \delta_{n_1} +\sigma_8 \delta_{m_1}  \right) u_1^{(0)} +
\sigma_9 \bar u_1^{(1)} u_2^{(2)}. \label{nls2}
\eeq
Then from eq.~(\ref{jh}) we see that~$u_2^{(1)}$ will satisfy the equation
$
\left( \sigma_1 \delta_{n_1} +\sigma_2 \delta_{m_1} \right) u_2^{(1)} = 0,
$
i.e. $u_2^{(1)}(n_1,\{m_i\}_{i=1}^K)=u_2^{(1)}(n_2,\{m_i\}_{i=2}^K)$ whenever eq.~(\ref{MMw2})
holds. Using eqs.~(\ref{nn},~\ref{equ22b}) we find that eq.~(\ref{nls2}) is equivalent to the
dNLS equation
\beq
\ri \delta _{m_2} u_1^{(1)} = \rho_1 \delta_{n_2}^2u_1^{(1)}+ \rho_2  u_1^{(1)} |u_1^{(1)}|^2, \label{nlsr}
\eeq
where
\bea
&& \rho_1 \doteq \frac{\ri \zeta \mu S^2 e^{\ri \kappa} (\zeta^2-\mu^2) \left( e^{2\ri \kappa} -1\right) }
{2 \tilde M_2 \left(\mu e^{\ri \kappa} -\zeta \right)}, \nonumber \\
&& \rho_2 \doteq \frac{\ri \zeta\mu (\mu-\zeta) \left(e^{2\ri \kappa} -1 \right)\left(e^{\ri \kappa} +1 \right)^4}
{\tilde M_2 e^{\ri \kappa} (\mu+\zeta) \left(\mu e^{\ri \kappa} -\zeta \right)^2  \left(\zeta e^{\ri \kappa} -\mu \right)^2}. \nonumber
\eea
Taking into account the form of~$S$,  one finds
that~$\rho_1$ and~$\rho_2$ are both real numbers:
\bea
&& \rho_1 = \frac{\mu \zeta r^2 (\zeta^2-\mu^2) \sin \kappa}
{ \tilde M_2 \left( \zeta^2 +\mu^2 -2 \zeta \mu \cos \kappa \right)}, \nonumber \\
&& \rho_2= -\frac{8 \zeta \mu (\zeta -\mu) (1 + \cos \kappa)^2 \sin \kappa}
{\tilde M_2(\mu+\zeta) \left(\zeta^2+\mu^2 -2  \zeta \mu \cos \kappa  \right)^2 }. \nonumber
\eea
Notice that~$\rho_1 \rho_2 <0$, so that eq.~(\ref{nlsr}) is a defocusing dNLS equation.

Eq.~(\ref{nlsr}) may be  obtained as a compatibility condition of the 
linear problem 
\bea 
&& \delta_{n_2} \Phi_{n_2,m_2}(\eta) = U(u_1^{(1)}, \bar u_1^{(1)}; \eta) \Phi_{n_2,m_2}(\eta),  \label{laxc1} \\
&& \delta_{m_2}\Phi_{n_2,m_2}(\eta)= V(u_1^{(1)}, \bar u_1^{(1)}; \eta) \Phi_{n_2,m_2}(\eta) , \label{laxc2} 
\eea
with
\bea \nonumber
 U(u_1^{(1)}, \bar u_1^{(1)}; \eta) &\doteq&\left(\begin{array} {cc}
\ri \eta & u_1^{(1)} \\
\bar u_1^{(1)}  & -\ri \eta  \end{array}\right),\qquad 
\\ \nonumber
 V(u_1^{(1)}, \bar u_1^{(1)}; \eta) &\doteq& \left(\begin{array} {cc}
2\ri \eta ^2+\ri |u_1^{(1)}|^2 & 2 \eta  u_1^{(1)} -\ri\delta_{n_2} u_1^{(1)} \\
2 \eta  \bar u_1^{(1)} +\ri \delta_{n_2} \bar u_1^{(1)}  & -2\ri \eta ^2-\ri |u_1^{(1)}|^2 \end{array}\right),
\eea
being~$\eta \in \mathbb{C}$ the spectral parameter and~$\Phi_{n_2,m_2}(\eta)$ a discrete complex vector function. 
A necessary and sufficient condition for the integrability of eq.~(\ref{nlsr}) is that 
the coefficients~$\rho_1$ and~$\rho_2$ be real. Thus eq.~(\ref{nlsr}) is integrable iff~$\ell =\infty$.

\section{Concluding remarks} \label{sec4}
In this paper we have extended the results obtained in~\cite{levi,lp} to the case of functions of slow-varyness of order infinity. 
To do so we have used the connection between shift operators and infinite series of differential operators (\ref{tnr}). 
Using these formulas we have been able to easily recover the formula (\ref{difInt}), 
usually proved by combinatorial techniques~\cite{Jordan}. Moreover, in analogy with the continuous case we can 
 introduce partial shift operators, which are expressed in terms of~$\delta$ operators, namely infinite series of difference operators. 

The multiscale expansion of the lpKdV equation in terms of harmonics implies, as a condition for the non existence of secular terms, a NLS equation written in terms of~$\delta$ operators. Choosing~$\delta_{n_2}$ in eq.~(\ref{nlsr}) as a series of symmetric differences and fixing
$\ell = 2$ the resulting dNLS equation is equivalent 
to the one presented in~\cite{levi,lp}, see eq.~(\ref{dnlstt}). 
In this way we have shown that for any finite order of slow-varyness the obtained dNLS equation  
will be local and no summation terms will ever appear. 

As shown in  Section 3, eq.~(\ref{nlsr}) has a matrix Lax pair (\ref{laxc1},~\ref{laxc2}), which is also expressed in terms of~$\delta$ operators. Such Lax pair has a reduction to any finite order  of slow-varyness, provided that the associated wave functions~$\Phi_{n_2,m_2}(\eta)$ have the same finite order  of slow-varyness.  However in the case of finite~$\ell$ the Lax pair~(\ref{laxc1},~\ref{laxc2}) reduces to difference equations and their compatibility is no more satisfied by the slow-varying approximation of eq.~(\ref{nlsr}). 
In fact the difference operators~$\Delta$ do not satisfy the Leibniz rule. 

This work still leaves many open problems on which we are working at the moment. Let us just mention the analysis of the solutions of the lpKdV equation obtained  from the exact ones of the reduced equation and their comparison with the ones obtained by carrying out a multiscale expansion of the continuous potential KdV equation; the reduction of symmetries of the lpKdV equation; the reduction of differential-difference equations both integrable and nonintegrable. 

It is still an open problem to understand the role played by an integrable discrete equation like the
nonlocal discrete NLS equation introduced by Ablowitz and Ladik~\cite{al} from the isospectral compatibility of a discrete analogue of the Zakharov-Shabat spectral problem~(\ref{laxc1},~\ref{laxc2}):
\bea \label{ddNLS}
\ri \frac{\Delta_m u_{n,m}}{\Delta t}& =& \frac{1}{2 (\Delta x)^2} \bigg[  \bigg( u_{n+1,m} - 2 u_{n,m} + u_{n-1,m} \prod_{k=-\infty}^{n-1} \Lambda_{k,m} \bigg) + \\ \nonumber
& +&\bigg( u_{n+1,m+1} \prod_{k=-\infty}^{n} \Lambda_{k,m} - 2 u_{n,m+1} + u_{n-1,m+1} \bigg) \bigg] \pm \nonumber \\
& \pm & \frac{1}{4} \bigg[ u_{n,m} \left( \bar u_{n,m} u_{n+1,m} +  \bar u_{n,m+1} u_{n+1,m+1} \right)  + \nonumber \\
&+ & u_{n,m+1} \left( u_{n-1,m} \bar u_{n,m} + u_{n-1,m+1} \bar u_{n,m+1} \right)  + \nonumber \\ \nonumber
&+& 2 |u_{n,m}|^2 u_{n+1,m+1} \prod_{k=-\infty}^{n} \Lambda_{k,m} + 2 |u_{n,m+1}|^2 u_{n-1,m} \prod_{k=-\infty}^{n-1} \Lambda_{k,m} \bigg] -\\ \nonumber
&-& u_{n,m} \sum_{k=-\infty}^{n} \Delta_m S_{k,m} - u_{n,m+1} \sum_{k=-\infty}^{n-1} \Delta_m \bar S_{k,m} ,
\eea
with
$$
\Lambda_{k,m} \doteq \frac{1 \pm |u_{k,m+1}|^2}{1 \pm |u_{k,m}|^2}, \qquad \qquad
S_{k,m} \doteq u_{k,m} \bar u_{k-1,m} + u_{k+1,m} \bar u_{k,m}.
$$
Eq.~(\ref{ddNLS}) reduces to the continuous NLS
equation when~$\Delta t \rightarrow 0$ and~$\Delta x \rightarrow 0$

\section*{Acknowledgments}
RHH was partially supported by the Region of Madrid and Universidad Polit\'ecnica de Madrid (UPM) through the grant ref.~CCG06-UPM/MTM-539. DL and MP were partially supported by PRIN Project ÔSINTESI-2004Õ of the Italian Minister
for Education and Scientific Research.  MP was partially supported by
the European Community through the FP6 Marie Curie RTN ENIGMA (contract number
MRTN-CT-2004-5652).


\end{document}